\documentclass[useAMS]{mn2e}
\def\mic {\hbox{\,$\umu$m }}
\def\IRAS{\textit{IRAS}}

\usepackage{amsmath, amssymb}
\usepackage{graphicx}
\usepackage{subfigure} 
\usepackage{times}

\begin{document}
\title[The FIR-Radio Relationship at High and Low Redshift]
{The FIR-Radio Relationship at High and Low Redshift}
\author[Catherine Vlahakis, Stephen Eales and Loretta Dunne]
{Catherine Vlahakis$^1$\thanks{E-mail: vlahakis@astro.uni-bonn.de; eales@astro.cf.ac.uk},
Stephen Eales$^2$\footnotemark[1]
and Loretta Dunne$^3$\\
$^1$Argelander-Institut f\"ur Astronomie\thanks{Founded by merging of the
Sternwarte, Radioastronomisches Institut and Institut f\"ur Astrophysik
und Extraterrestrische Forschung der Universit\"at Bonn}, Universit\"at Bonn, 
Auf dem H\"ugel 71, 53121 Bonn, Germany\\
$^2$School of Physics and Astronomy, Cardiff University,
\\ 5, The Parade, Cardiff CF24 3AA\\
$^3$School of Physics and Astronomy, University of Nottingham, 
\\University Park, Nottingham, NG7 2RD\\}
\maketitle

\begin{abstract}
We use the results of the SCUBA Local Universe Galaxy Survey, a 
submillimetre survey of galaxies in the nearby Universe, to
investigate the
relationship between the 
far-infrared--submillimetre and radio emission of galaxies
at both low and high redshift. At low redshift we show that
the correlation between radio and far-infrared emission is much
stronger than the correlation between radio and submillimetre
emission, which is evidence that massive stars are the source
of both the far-infrared and radio emission. At high redshift we show
that the submillimetre sources detected by SCUBA are brighter
sources of radio emission than are predicted from the properties of
galaxies in the local Universe. We discuss possible reasons for the
cosmic evolution of the relationship between radio and far-infrared emission.

\end{abstract}


\section{Introduction}
\label{intro}

One of the most intriguing relationships in astronomy is that
between the non-thermal radio and far-infrared (FIR) emission from galaxies.
It is one of the strongest correlations in astronomy, with a tight correlation between 
FIR and radio luminosity over five decades of luminosity (e.g. Price \& Duric 1992; Yun, 
Reddy \& Condon 2001). The relationship holds for a remarkably wide variety of galaxy 
types, and is observed not only globally but also on $\sim$kiloparsec 
scales within the disks of individual galaxies (e.g. Boulanger \& 
P\'erault 1988; Bicay \& Helou 1990; Fitt et al. 1992; Murphy et al. 2006). Some authors 
also find evidence of the relationship at high redshifts (e.g. Garrett 2002; Appleton et al. 
2004; Kov\'acs et al. 2006). Yet, although the 
relationship has been known for over two decades (e.g. Condon 1992 and references 
therein), the cause of the relationship is still unclear.

The standard explanation of the relationship is that both the
FIR and radio emission are caused
by high-mass stars, with the stars both heating the dust, which then emits
FIR emission, and producing the relativistic electrons
responsible for synchrotron radiation when
at the end of their lives they explode as supernovae (see e.g. Condon 1992 and references 
therein).
Alternative explanations have been proposed. However, most workers
in the field have assumed that the standard explanation (i.e. ongoing star-formation) is 
the correct one.

Nevertheless, despite its widespread acceptance, there is surprisingly little
direct evidence for the standard explanation and it has one major problem. The problem is 
that the slope of the FIR--radio relation is
roughly unity (Helou, Soifer \& Rowan-Robinson 1985;
Devereux \& Eales 1989) whereas the synchrotron emissivity does not just depend on
the number-density of relativistic electrons but also on the strength of
the magnetic field.
Although there are only a few measurements
from Zeeman splitting of the magnetic fields in galaxies (Thompson et al.
2006 and references therein), it is possible to estimate the magnetic
field in a galaxy using the `minimum energy' argument (Longair 1983), and
estimates of the magnetic field made in this way can vary by as much as a factor
of 100 from normal galaxies like our own to vigorously star-forming galaxies
such as Arp 220 (Thompson et al. 2006). The radio spectrum of a galaxy
roughly follows a power-law ($S \propto \nu^{-\alpha}$) and the
synchrotron emissivity is proportional to $B^{1.0+\alpha}$. The radio emissivity in Arp 220 
will therefore be much higher than in our own galaxy, not only because of the much greater 
production rate of relativistic electrons per cubic kiloparsec, but also because of the 
stronger magnetic field. However, it is still possible to explain
 the unity slope of the FIR--radio relation if the 
relativistic electrons lose all their energy within their galaxy, and thus
an electron in Arp 220 loses its energy at a faster rate but ultimately only loses the
same amount of energy as in a galaxy like our own. 

This scenario is described by the `calorimeter' model (V\"olk 1989; 
Lisenfeld, V\"olk \& Xu 1996), in which the dust-heating stellar UV photons 
and the relativistic electrons lose most of their energy within the galaxy. The 
calorimeter model will only give a unity slope for the FIR--radio relation if the ratio 
of inverse Compton to synchrotron losses is roughly the same in all galaxies, which leads 
to the prediction that the energy density of the interstellar radiation field and that 
of the magnetic field have a constant ratio for galaxies. Lisenfeld et al. (1996) used 
observations of 114 `normal' late-type galaxies to 
estimate the mean energy densities of the interstellar radiation field and, using the 
`minimum energy' argument, the magnetic field. They showed a linear correlation between 
the two, suggesting turbulent dynamo activity in galaxies as a possible explanation. 
Calorimeter-type models have also been 
presented by several other authors (e.g. Pohl 1994). 

A more fundamental problem with the calorimeter model is that the radio spectra of
star-forming galaxies do not exhibit the steepening that one would expect
if the electrons lose all their energy within the galaxy in which they
were formed. A recent interesting suggestion has been that the
magnetic fields in galaxies are actually much higher than the estimates
from the minimum energy argument, which would definitely lead to an electron losing
all its energy before it diffuses out of its galaxy (Thompson et al. 2006).
Thompson et al. explain the lack of clear curvature in the radio spectra as
the result of ionization losses, which flatten the spectrum at low 
frequencies.

An alternative idea is that the FIR--radio relation holds for the `optically thin' 
scenario in which the majority of the stellar UV photons and relativistic electrons do escape 
from the galaxy disk (e.g. Helou \& Bicay 1993). The FIR--radio correlation is then 
attributed to a constant ratio between the production rates of dust-heating UV photons and 
synchrotron-emitting relativistic electrons.

Niklas \& Beck (1997) argue that observations do not support either the calorimeter or 
`optically-thin' model and suggest an alternative model in which the FIR--radio relation 
arises from a combination of correlations; that  between the volume density of cool 
(neutral) gas and the star formation rate and magnetic field strength, and that between 
the star formation 
rate and the FIR luminosity, assuming equipartion between the cosmic ray energy density and 
the magnetic field strength. In this scenario the cool neutral gas clouds play a central 
role. Such a model, unlike either the calorimeter (optically thick) or optically thin 
models, can explain the fact that the FIR--radio relation is not only observed globally 
but is also observed on kiloparsec scales within the disks of individual galaxies.

Alternative explanations have also been proposed by a number of other authors. Bettens et al.
(1993) proposed that the cosmic ray flux, 
which as 
in the `standard explanation' produces the non-thermal radio emission, drives the 
evolution of molecular clouds; a higher flux of cosmic rays produces a higher
ionization fraction in the cloud, which delays the cloud's collapse, and
ultimately leads to a larger
fraction of OB stars and more FIR emission from the
surrounding dust. However, other authors (e.g. Niklas \& Beck 1997, Murgia et al. 2005) 
have pointed out problems with this 
model; for example it fails to take into account the effect of the the interstellar 
magnetic field strength on the synchrotron luminosity. 

Murgia et al. (2005) proposed a model in which the relation between CO, radio continuum and 
FIR emission results from hydrostatic pressure in galaxies, which can explain the 
correlation on both global and local scales. In this model the relationship between the 
CO and radio is explained by hydrostatic pressure acting as a regulating mechanism 
between the molecular gas fraction and the magnetic field strength, while the link to 
the FIR emission comes from models by Dopita et al. (2004) which connect FIR emission 
with interstellar pressure. This model thus avoids an explicit dependence on star formation.

Most of the various models proposed to explain the FIR--radio relationship require a link 
between the gas density and the magnetic field strength. One suggestion as to possible 
physical processes behind such a model is magnetohydrodynamic turbulence (Groves et al. 
2003).

Whatever its cause, the strong correlation between FIR and radio emission in nearby
galaxies provided submillimetre (submm) astronomers with a valuable tool for estimating the
redshifts of the galaxies discovered in the
first deep submm surveys
(Smail, Ivison \& Blain 1997; Hughes et al. 1998; Eales et al. 1999).
These galaxies are very faint at optical wavelengths, which has
made it very difficult to measure redshifts using conventional
spectroscopy, and until recently
very few spectroscopic redshifts existed (Chapman et al. 2005).
Carilli \& Yun (1999) first suggested that an alternative to optical
spectroscopy would be to 
assume that the FIR--radio relation is the same at all redshifts, and
then
estimate redshifts from the ratio of submm
to radio flux densities.

In this paper we explore the FIR-radio relationship at both high and low redshifts,
using the results of the recently-completed
SCUBA Local Universe Galaxy Survey (SLUGS). This is a submm survey
of $\sim 200$ galaxies in the local Universe, half drawn from a sample selected
in the FIR (Dunne et al. 2000b; herein the IRS sample) and half drawn from 
an optically-selected (herein OS) sample
(Vlahakis, Dunne \& Eales 2005). Our submm measurements allow us to test
a basic prediction of the standard theory. Because of the greater intensity
of the interstellar radiation field in regions containing large
numbers of OB stars, the dust in these regions 
will be hotter than the dust in the general interstellar medium. Therefore,
the standard hypothesis makes the prediction that the correlation between FIR and
radio emission will be tighter than the correlation between submm and radio
emission. Previous authors have found evidence both in favour of and against this 
prediction by decomposing the FIR emission into warm and cold components (e.g. Hoernes, 
Berkhuijsen and Xu 1998; Pierini et al. 2003), but this previous work was based on FIR 
measurements alone rather than measurements at submm wavelengths. In Section~\ref{fir-radio} 
we use our SLUGS data to test the prediction. 
In Section~\ref{indicator} we then consider the FIR--radio relationship at high redshift. 
Because SLUGS is the
only large sample of galaxies for which there are measured spectral energy distributions
extending through the FIR and submm wavebands, we can predict the ratios
of radio to submm emission one would see at high redshift {\it if there is no 
cosmic evolution in the FIR--radio relationship.} Dunne, Clements \& Eales (2000) 
performed a similar analysis using the IRS data alone. Using the spectroscopic 
redshifts from Chapman et al. (2005), we show that the sources detected in deep SCUBA 
surveys are generally brighter radio sources than one would predict from the FIR--radio 
relationship seen in the local Universe.   

\section{The FIR--radio relationship in the low-redshift Universe}
\label{fir-radio}

\begin{table}
\centering
\begin{minipage}[t]{8cm}
\caption{\label{radio-tab}\small{Literature 1.4\,GHz Radio fluxes.}}
\begin{tabular}{lrrr}
\\[0.5ex]
\hline \\[-2ex]
(1) & (2) & (3) & (4) \\
\\[-2ex]
Name & $S_{1.4}$ & log $L_{1.4}$ & $\alpha^{850}_{1.4}$\\
& (m\,Jy) & \small{(W\,Hz$^{-1}$sr$^{-1}$)} & \\
\\[-1.0ex]
\hline 
\\[-1.8ex]
UGC 148 & 17.8 & 20.71 & 0.21 \\
NGC 99 & 8.2 & 20.58 & 0.37 \\
PGC 3563 & 4.1 & 20.31 & 0.34 \\
NGC 786 & 6.1 & 20.31 & 0.43 \\
NGC 803 & $<$1.5 & $<$19.02 & $>$0.75 \\
UGC 5129 & 1.8 & 19.68 & $<$0.54\\
NGC 2954 & $<$0.84 & $<$19.30 & ... \\ 
UGC 5342 & 3.8 & 20.11 & 0.39 \\
PGC 29536 & 1.0 & 20.15 & $<$0.68 \\
NGC 3209 & 1.11 & 19.84 & $<$0.54 \\
NGC 3270 & 6.2 & 20.60 & 0.41 \\
NGC 3323 & 13.4 & 20.76 & 0.30 \\
NGC 3689 & 27.1 & 20.52 & 0.24 \\
UGC 6496 &  1.5 & 19.98 & 0.47 \\
PGC 35952 & 3.8 & 19.99 & 0.47 \\
NGC 3799/3800 & 49.3 & 20.94 & 0.18 \\
NGC 3812 & 1.96 & 19.62 & $<$0.54 \\
NGC 3815 & 3.2 & 19.85 & 0.47 \\
NGC 3920 & 7.6 & 20.21 & 0.27 \\
NGC 3987 & 56.3 & 21.27 & 0.22 \\
NGC 3997 & 5.2 & 20.28 & $<$0.27 \\
NGC 4005 & 2.4 & 19.89 & $<$0.33 \\
NGC 4015 & $<$1.5 & $<$19.66 & ... \\
UGC 7115 & 87.3 & 21.82 & $-$0.10\\
UGC 7157 & $<$1.4 & $<$19.92 & ...\\
IC 797 & 11.5 & 19.91 & 0.36 \\
IC 800 & 1.4 & 19.09 & 0.73 \\
NGC 4712 & $<$1.2 & $<$19.57 & $>$0.81 \\
PGC 47122 & $<$1.8 & $<$20.17 & ... \\
MRK 1365 & 23.0 & 21.06 & 0.06 \\
UGC 8872 & $<$1.1 & $<$19.74 & ... \\
UGC 8883 & 3.0 & 20.18 & $<$0.47 \\
UGC 8902 & 12.4 & 21.08 & 0.31 \\
IC 979 & $<$1.5 & $<$20.17 & $>$0.66 \\
UGC 9110 & 7.3 & 20.41 & $<$0.34 \\
NGC 5522 & 17.6 & 20.78 & 0.26 \\
NGC 5953/4 & 91.4 & 20.76 & 0.20 \\
NGC 5980 & 25.3 & 20.84 & 0.42 \\
IC 1174 & $<$1.5 & $<$19.73 & $>$0.51 \\
UGC 10200 & 11.2 & 19.85 & $<$0.11 \\
UGC 10205 & 2.9 & 20.31 & 0.55 \\
NGC 6090 & 48.0 & 21.79 & 0.12 \\
NGC 6103 & 5.9 & 20.94 & 0.40\\
NGC 6104 & 6.4 & 20.87 & $<$0.30 \\
IC 1211 & $<$1.3 & $<$19.83 & $>$0.56 \\
UGC 10325 & 11.2 & 20.77 & 0.24 \\
NGC 6127 & $<$1.2 & $<$19.66 & $>$0.78 \\
NGC 6120 & 33.7 & 21.67 & 0.12 \\
NGC 6126 & $<$1.5 & $<$20.37 & $>$0.50 \\
NGC 6131 & 6.5 & 20.44 & 0.39 \\
NGC 6137 & 438.7 & 22.80 & $-$0.50 \\
NGC 6146 & 162.7 & 22.32 & $-$0.32 \\
NGC 6154 & 1.2 & 19.85 & $<$0.64 \\
NGC 6155 & 13.9 & 20.12 & 0.08 \\
UGC 10407 & 18.4 & 21.33 & 0.06 \\
NGC 6166 & 3713.0 & 23.70 & $-$0.72 \\
NGC 6173 & 6.9 & 20.94 & $<$0.23 \\
NGC 6189 & 7.9 & 20.61 & 0.40 \\
NGC 6190 & 4.8 & 19.94 & 0.55 \\
NGC 6185 & 62.1 & 22.04 & $<-$0.13 \\
\\[-1.3ex]
\end{tabular} 
\end{minipage}
\end{table}

\begin{table}
\centering
\begin{minipage}[t]{7cm}
\contcaption{}
\begin{tabular}{lrrr}
\\[0.5ex]
\hline \\[-2ex]
(1) & (2) & (3) & (4)\\
\\[-2ex]
Name & $S_{1.4}$ & log $L_{1.4}$ & $\alpha^{850}_{1.4}$ \\
& (m\,Jy) & \small{(W\,Hz$^{-1}$sr$^{-1}$)} & \\
\\[-1.0ex]
\hline 
\\[-1.8ex]
UGC 10486 & $<$1.5 & $<$19.96 & ... \\
NGC 6196 & 1.0 & 20.17 & $<$0.57 \\
UGC 10500 & $<$1.4 & $<$19.79 & ... \\
IC 5090 & 32.6 & 21.67 & 0.23 \\
IC 1368 & 23.7 & 20.80 & 0.12 \\
NGC 7047 & 3.9 & 20.31 & 0.48 \\
NGC 7081 & 18.2 & 20.50 & 0.16 \\
NGC 7280 & $<$1.2 & $<$18.82 & ... \\
NGC 7442 & 7.8 & 20.83 & 0.32\\
NGC 7448 & 81.5 & 20.80 & 0.16 \\
NGC 7461 & $<$1.4 & $<$19.62 & ... \\
NGC 7463 & 16.3 & 20.16 \\
III ZW 093 & 5.3 & 21.30 & $<$0.29 \\
III ZW 095 & 1.5 & 20.14 & $<$0.46 \\
UGC 12519 & 4.6 & 20.16 & 0.51 \\
NGC 7653 & 92.2 & 21.43 & 0.04 \\
NGC 7691 & 4.5 & 20.08 & $<$0.31 \\
NGC 7711 & $<$1.2 & $<$19.51 & ...\\
NGC 7722 & 4.7 & 20.09 & 0.47 \\
\\[-1.8ex] \hline \hline
\\[-1.0ex]
\end{tabular}\\ 
\begin{small}
(1) Galaxy name; (2) 1.4\,GHz radio flux taken from the literature; NVSS (most) or FIRST; 
(3) 1.4\,GHz luminosity; (4) radio--submm spectral index ($\alpha^{850}_{1.4}$), 
calculated as described 
in Section~\ref{indicator}.
\end{small}
\end{minipage}
\end{table} 

The radio parameters for the OS sample are given in Table~\ref{radio-tab}. Radio 
(1.4\,GHz) fluxes were taken from the literature, most from the NRAO/VLA Sky Survey 
(NVSS; Condon et al. 1998) or 
otherwise from the VLA FIRST (Faint Images of the Radio Sky at Twenty Centimeters) 
survey (Becker, White \& Helfand 1995). For objects without literature radio fluxes 
the NVSS radio images were used 
to either measure the flux of a faint source or, where no source was seen, to  
measure an upper limit. The 1.4\,GHz luminosities were calculated using the standard 
equation (e.g. Condon et al. 1990). Apart from a few 
measured radio spectral 
indices we usually assumed a value of 0.7 (Condon 1992). For the IRS sample, radio 
parameters are those used in Dunne et al. (2000a), which were taken from Condon et 
al. (1990); where 
there were flux measurements in the literature at more than one wavelength spectral 
indices were calculated, and 
otherwise a value of 0.7 was assumed.

\begin{figure}
\begin{center}
\includegraphics[angle=0, height=6cm]{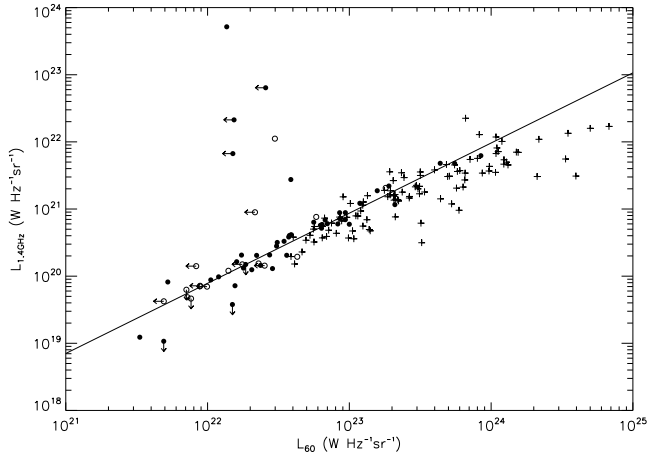}
\caption{\label{lum-radio-lor}{\small 1.4\,GHz luminosity versus 60\mic luminosity for 
the OS and IRS samples (circles and crosses respectively; for the OS galaxies filled 
circles indicate those detected at 850\mic while open circles indicate the non-detections 
at 850\,\micron\/ (see Vlahakis et al. (2005)); for clarity, OS galaxies with both 60\mic 
and radio upper limits are not shown). The 
solid line shows a linear least-squares fit to the OS galaxies only, where the slope 
of the log-log plot is 1.04$\pm$0.04. The 7 OS outliers which have `excess' 
1.4\,GHz emission relative to this correlation are all early-type galaxies.}}
\end{center}
\end{figure}

\begin{figure}
\begin{center}
\includegraphics[angle=0, height=6cm]{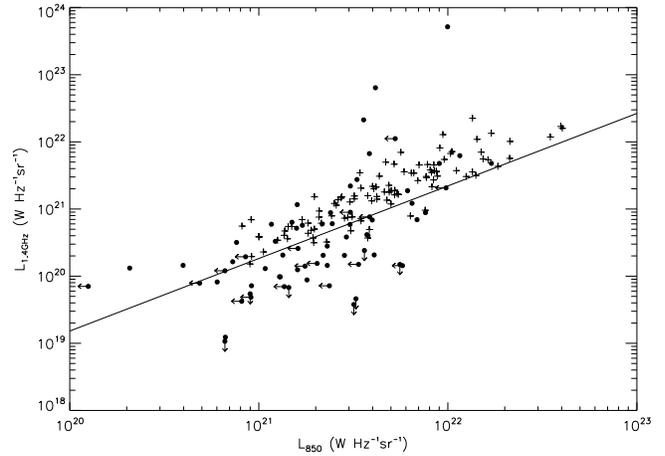}
\caption{\label{lum850-radio-lor}{\small 1.4\,GHz luminosity versus 850\mic luminosity 
for the OS and IRS samples (filled circles and crosses respectively; for clarity, OS 
galaxies with both submm and radio upper limits are not shown). The solid line 
shows a linear fit to
 the OS data alone, where the slope of the log-log plot is 1.08$\pm$0.16. 
There is a correlation between radio 
and 850\mic luminosity for the OS sample \mbox{($r_{s}$\,=\,0.71, probability$<$0.001)} but 
there is much more scatter than for the relation with 60\mic luminosity.}}
\end{center}
\end{figure}

The relationship between 1.4\,GHz luminosity and 60\mic luminosity is shown in 
Figure~\ref{lum-radio-lor}. The filled points indicate OS objects detected at 850\mic 
while for comparison the open points represent the OS non-detections at 850\,\mic 
(see Vlahakis et al. 
2005). The IRS galaxies are plotted as cross symbols. In the following analysis only 
OS galaxies with detections at the relevant wavelengths and not upper limits are 
considered. For the OS sample there
 is a very tight correlation between 1.4\,GHz luminosity and 60\mic luminosity 
(correlation coefficient $r_{s}$=\,0.97, 
probability$<$0.001), with a less tight correlation for the IRS sample. 
Figure~\ref{lum-radio-lor} also shows the linear least-squares fit to the OS sample 
alone (solid line), where the slope of the log-log plot is 1.04$\pm$0.04. We note that 
the slope for the OS sample, both for Figure~\ref{lum-radio-lor} and 
Figure~\ref{lum850-radio-lor}, was calculated using only those galaxies with detections and 
not upper limits, and thus 
this may affect the exact value of the slope. Apart from this, any apparent 
discrepancy between our value of the slope and that found by previous authors 
(e.g. Cox et al. 1988 and Devereux \& Eales 1989) 
could also be due to the fact that a) we plot 60\mic luminosity whereas the previous 
authors used the FIR luminosity, and b) our SLUGS sample includes higher 60\mic 
and FIR luminosities than these previous studies.

The detected OS galaxies appear to follow 
an even tighter 60\,\micron--radio relation than the IRS galaxies. The reason for 
this is not clear, but it could be related to the fact that the luminous \IRAS\/ 
galaxies tend to be mergers or more active systems. Another possible explanation 
could be that the several objects in the IRS sample which appear to be under-luminous 
in the radio (including NGC\,4418; see discussion by Roussel et al. 2003) are undergoing a 
very recent burst of intense star-formation and so there has not been time for the number 
of relativistic electrons to reach the equilibrium value. We note, however, that the 
apparent difference in scatter may also be explained by the large number of upper limits for 
the OS galaxies.

There are 7 extreme outliers (all OS galaxies) which have excess 1.4\,GHz emission 
relative to this correlation, all of which are early-types (the furthest outlier 
is NGC\,6166). The `excess' 1.4\,GHz emission most likely indicates the presence 
of an AGN in these galaxies. 

The relationship between 1.4\,GHz luminosity and 850\mic luminosity is shown in 
Figure~\ref{lum850-radio-lor}. For the OS sample there is a correlation ($r_{s}$=\,0.73, 
probability$<$0.001), but there is clearly a much tighter correlation between radio and 
60-\micron\/ luminosity than between radio and 850-\micron\/ luminosity (the rms 
scatter is 0.13 for 60\,\micron--radio and 0.40 for 850\,\micron--radio). 
This is exactly the behaviour we would 
expect if the standard model is correct (see Section~\ref{intro}). 
We would expect to see a stronger correlation between the 60-\micron\/ and radio 
luminosity, because the 60-\micron\/ emission is dominated by emission from hot dust 
in star-formation 
regions mainly heated by OB stars, whereas the 850-\micron\/ emission traces 
colder dust, which is heated by the general 
interstellar radiation field, which must include a component from older stellar 
populations. Vlahakis et al. (2005) showed that the OS galaxies contain large 
proportions of cold dust, with up to three orders of magnitude more mass of cold dust 
than warm dust. The fact that a much larger 
scatter at 850\mic is precisely what we see thus provides evidence that the 
standard explanation for the FIR--radio luminosity correlation is correct and 
that it is the population of young (OB) 
stars which both heats the dust and provides the source of radio emission. 

Comparing the OS galaxies to the IRS galaxies there are some interesting points to 
note from Figure~\ref{lum850-radio-lor}. i) Both the OS and IRS samples follow a 
similar slope (the line in Figure~\ref{lum850-radio-lor} shows the linear fit 
for the OS sample alone, where the slope of the log-log plot is 1.08$\pm$0.16), but while 
for the OS sample the scatter is  much greater than seen for the 60\micron--radio 
relation (with implications which are discussed 
above), for the IRS sample there is a 
similar amount of scatter for both 60-\micron\/ and 850-\micron\/ luminosity with the 
radio ($\sim$\,0.2). 
ii) There is a much larger scatter for the OS sample than the IRS sample (0.22 for the IRS 
sample and 0.40 for the OS sample, excluding those galaxies with `excess' radio emission 
which are likely AGN-related). We can explain both these facts if for the IRS galaxies 
warm dust dominates the 
emission at both 60- and 850-\micron\/ (see Vlahakis et al. 2005).

\section{The FIR--radio relationship in the high-redshift Universe}
\label{indicator}

Carilli \& Yun (1999) proposed a method of using the redshift-sensitive nature of the 
radio--submm flux density ratio as a redshift estimator. The flux density ratio $S_{850}/S_{1.4}$
 is related to the radio--submm spectral index between 1.4\,GHz and 850\mic 
($\alpha^{850}_{1.4}$) by the equation
\begin{equation}\label{eq:spec}
\alpha^{850}_{1.4}=0.42 \times \mathrm{log}\left(\frac{S_{850}}{S_{1.4}}\right)
\end{equation}
where $S_{850}$ and $S_{1.4}$ are the flux densities at 850\mic and 1.4\,GHz respectively 
(as defined in Carilli \& Yun 1999). Carilli \& Yun (2000) predicted 
the evolution of $\alpha^{850}_{1.4}$ with 
redshift for 17 low-redshift star-forming galaxies by making polynomial fits to 
the observed FIR--radio SEDs of these galaxies. Likewise, Dunne et al. (2000a) predicted 
the change of $\alpha^{850}_{1.4}$ with 
redshift for the 104 galaxies in the IRS SLUGS sample. This method is based on 
the assumption that the 
FIR--radio luminosity relation is the same at low and high redshifts, i.e. that 
more distant galaxies have similar properties (e.g. magnetic field strength or 
dust grain properties) to those observed in the local Universe. As discussed in 
Carilli \& Yun (2000), other factors which could potentially 
affect the reliability of this method include the spectral index assumed for the 
radio synchrotron emission (which typically has values of 0.7 to 0.8 for star-forming 
galaxies (Condon 1992)), and any dependence on galaxy properties such as dust 
temperature; Blain (1999) showed that this method would be unable to distinguish 
hot dusty objects at high redshift from 
cold dusty objects at low redshift. A final problem is the possible presence of 
an AGN in one of the high-redshift sources.

Previous attempts to predict how $\alpha^{850}_{1.4}$ should depend on redshift 
have used samples of low-redshift galaxies that are selected in the FIR. Our 
previous work on the OS SLUGS sample (Vlahakis et al. 2005) found a population 
of cold dusty galaxies (T $\sim$\,20\,K) which is missing from \IRAS\/-selected 
samples, and thus these `cold' galaxies have so far been 
unrepresented in any analysis of the effectiveness of the Carilli \& Yun method as 
a redshift indicator. The rest-frame wavelength for any source at moderate redshifts 
($z<2$) detected in the SCUBA surveys will be at a significantly longer wavelength 
than the \IRAS\/ bands, so a SCUBA galaxy is likely to contain more cold dust 
than a low-redshift galaxy detected by an \IRAS\/ survey. 

\begin{figure}
\begin{center}
\includegraphics[angle=0, width=8cm]{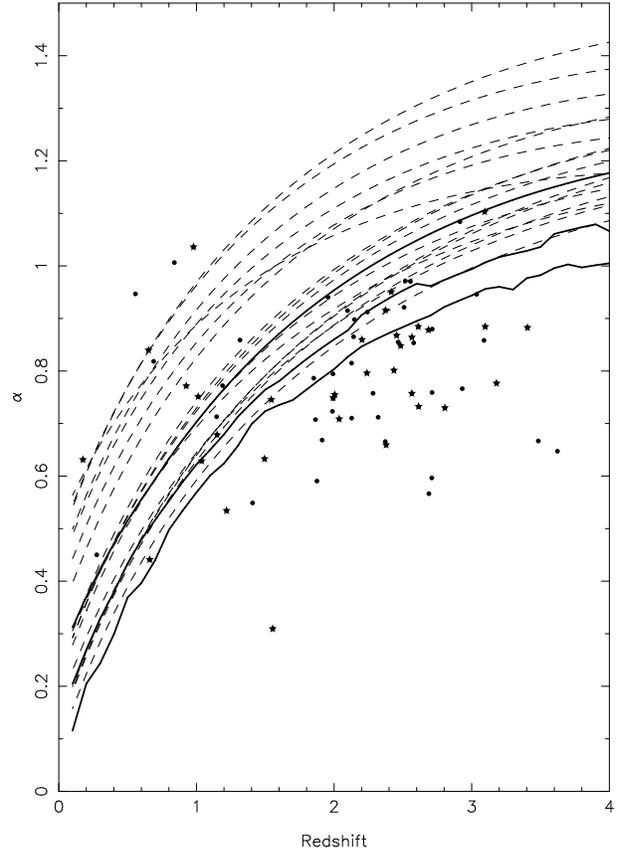}
\caption[The $\alpha^{850}_{1.4}$--z relationship for the 17 OS SLUGS galaxies 
detected at both 850- and 450-\,\micron.]{\label{high-red}{\small The 
$\alpha^{850}_{1.4}$--z relationship for the 17 OS SLUGS galaxies detected 
at both 850 and 450\,\micron, shown as the dashed lines. The 
thick solid lines show the median of the predictions for the 25 most luminous 
IRS sample galaxies, together 
with the $\pm1\sigma$ spread for these predictions. The plotted points in the diagram show 
all the deep SCUBA sources with both spectroscopic redshifts and radio measurements 
from Chapman et al. (2005). Stars indicate those which have the most robustly determined 
redshifts, following Aretxaga et al. (2007). The fact that the OS galaxies tend to have 
higher values of $\alpha^{850}_{1.4}$ than the IRS galaxies can clearly be seen, with 
about two thirds of the OS curves having higher $\alpha^{850}_{1.4}$ at a given 
redshift than the $1\sigma$ variation of the IRS predictions. }}
\end{center}
\end{figure}

In this section we use the OS SLUGS galaxies to make a new analysis of the 
$\alpha^{850}_{1.4}$--redshift relation, in a similar way to that carried 
out by Dunne et al. (2000a) for the IRS sample. We use 
the fitted SEDs for the 17 OS galaxies detected at both 850\mic and 450\mic 
(fitted using a two-component model to the 60\,\micron, 100\,\micron, 450\,\mic 
and 850\mic data points as described in Vlahakis et 
al. (2005)). The 1.4\,GHz fluxes used are listed in Table~\ref{radio-tab}. The change in 
$\alpha^{850}_{1.4}$ with redshift for each of the OS galaxy SEDs is plotted in 
Figure~\ref{high-red}, shown as the dashed lines. The thick solid lines show the 
median of the predictions for the 25 IRS galaxies with the highest 60-\mic 
luminosities, together with the $\pm1\sigma$ variation in the predictions. The plotted 
points in the diagram show all the deep SCUBA sources with 
both spectroscopic redshifts and radio measurements from Chapman et al. (2005). 
Stars indicate the Chapman et al. sources which have the most robustly determined 
redshifts --- following Aretxaga et al. (2007), those whose 
spectroscopic redshifts are derived from the identification of at least two spectral 
features and which have unambiguous optical/IR/radio counterparts.

\begin{figure}
\begin{center}
\subfigure[]{\label{alpha-850-both}
\includegraphics[angle=0, width=8.5cm]{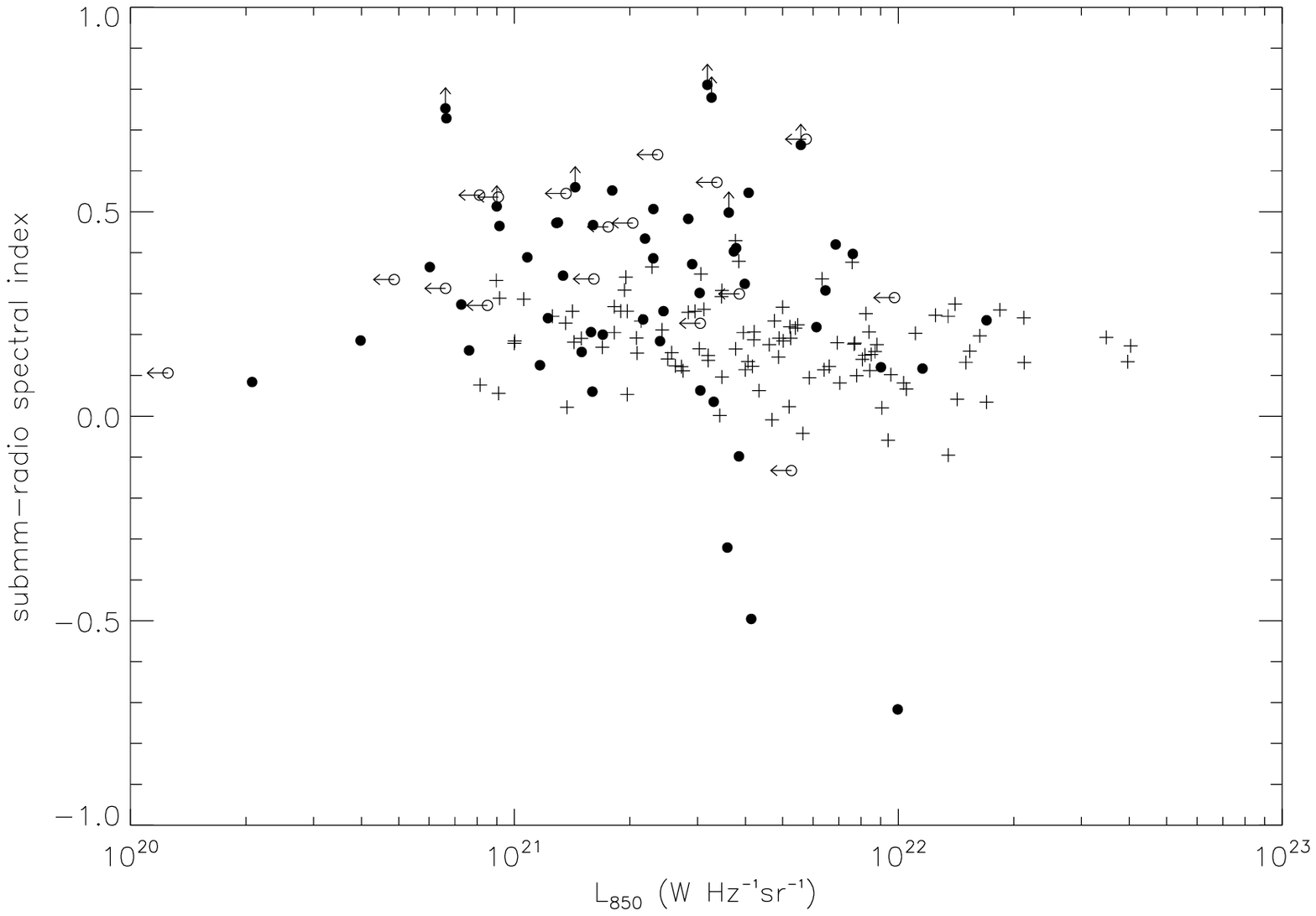}}
\subfigure[]{\label{alpha-s60-both}
\includegraphics[angle=0, width=8.5cm]{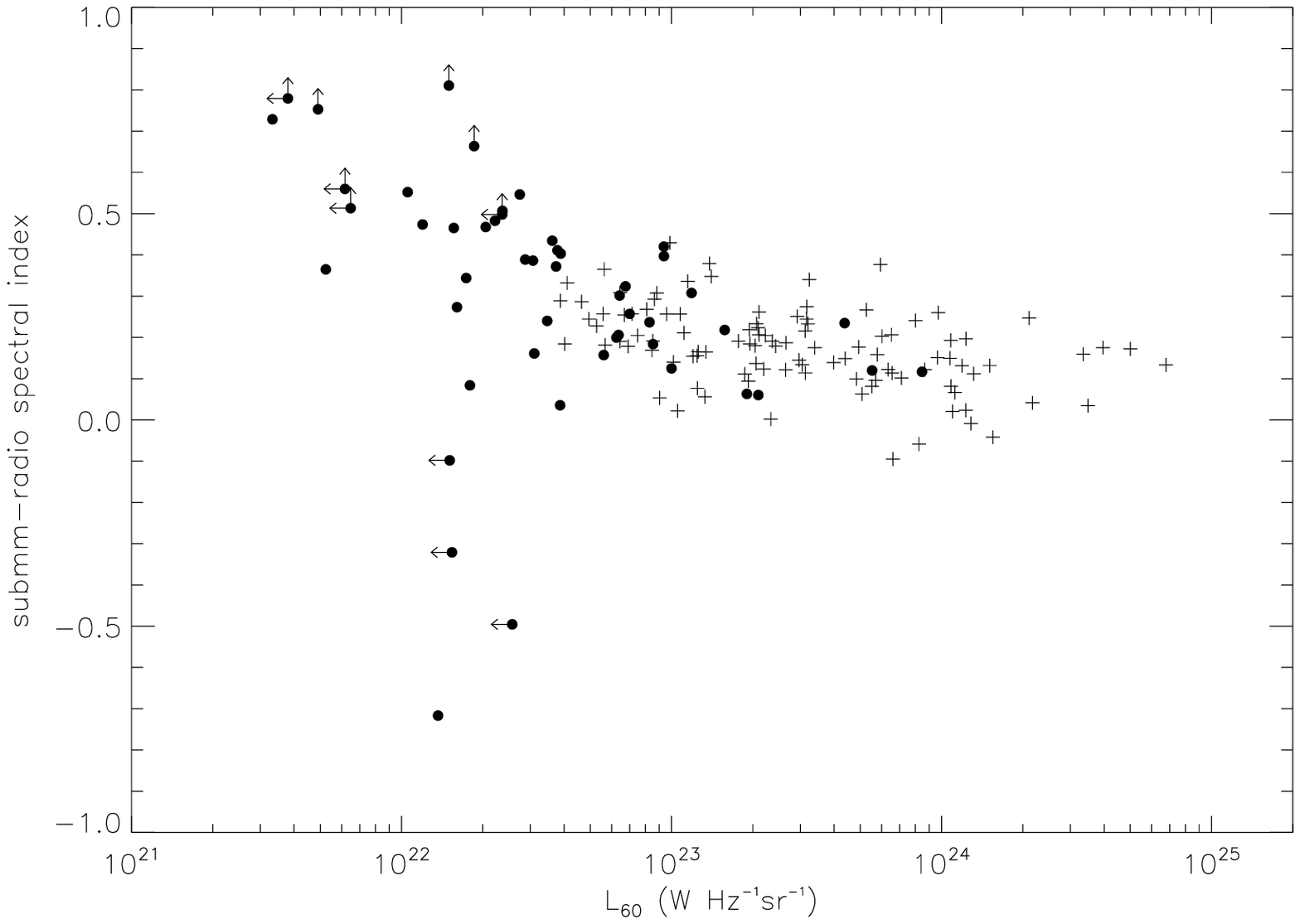}}
\caption{\label{alpha-plots}Submm--radio spectral index, $\alpha^{850}_{1.4}$, 
versus (a) 850-\mic luminosity and (b) 60-\micron\/ luminosity, for the OS and 
IRS samples. Filled circles indicate OS galaxies detected at 850\mic while open 
circles ((a) only) represent the non-detections at 850\,\micron; galaxies with 
both submm and radio upper limits are not plotted. The IRS galaxies are 
plotted as cross symbols.}
\end{center}
\end{figure}

It is immediately evident that many of the OS galaxies predict a very different 
$\alpha^{850}_{1.4}$--redshift relationship to the IRS galaxies, with about two 
thirds of the OS galaxies lying outside the $1\sigma$ variation of the IRS 
predictions and having higher $\alpha^{850}_{1.4}$ at a given redshift. Therefore, 
in the absence of an independent 
measurement of redshift, a given source 
could be either IRS-like and at higher redshift or OS-like and at lower redshift,
which is essentially the degeneracy noted by Blain (1999). 
When the OS galaxies are included, it can be seen that the uncertainty in the 
Carilli \& Yun method
is much greater than would be evident from the IRS curves alone.
In summary, it seems that the temperature of the bulk of the dust significantly affects the 
position 
of a galaxy on the $\alpha^{850}_{1.4}$--z diagram, and that as a consequence of 
the differing proportions of warm and cold dust in local galaxies, which is quite 
clearly shown in the 
comparison of the two SLUGS samples, the Carilli \& Yun technique is unlikely to 
be reliable as a redshift estimator for deep submm sources without measuring the 
temperature of the dust in these objects.

We now turn to the measurements for the high-redshift galaxies. The main thing to note from 
Figure~\ref{high-red} is 
that the high-redshift sources are significantly 
brighter at radio wavelengths, or alternatively fainter at FIR wavelengths, than 
the predictions for either the OS or IRS sample. Lewis, Chapman \& Helou (2005) find a 
similar result comparing their model predictions to measurements of deep submm sources.
Evidence that the ratio of radio to FIR emission may be higher at high redshift than at 
low redshift has also been found by other authors (e.g. Kov\'acs et al. 
2006, who used additional submm observations 
of a subset of the Chapman et al. (2005) sample).

There are a four possible explanations for the difference between the Chapman 
et al. sources and the SLUGS $\alpha^{850}_{1.4}$--z predictions. Only one is 
that there is a genuine difference between the
FIR--radio relationships at high and low redshift.

\begin{enumerate}

\item{The high-redshift SCUBA galaxies have an average dust luminosity that is 
about four times greater than the most luminous SLUGS galaxy (Coppin et al., 
in preparation). Therefore, if there is an inverse correlation between 
$\alpha^{850}_{1.4}$ and luminosity at zero redshift, 
our models will under-predict the radio emission of the high-redshift galaxies. 
We investigate this possibility in Figure~\ref{alpha-plots}. There is only a 
weak inverse correlation between $\alpha^{850}_{1.4}$ and 850-\mic luminosity 
(Figure~\ref{alpha-850-both}). There is an inverse 
correlation between $\alpha^{850}_{1.4}$ and 60-\mic luminosity 
(Figure~\ref{alpha-s60-both}), but it 
is unclear whether it continues at luminosities $\gtrsim10^{24}$ W\,Hz$^{-1}$sr$^{-1}$. 
The discrepancy in Figure~\ref{high-red} between our IRS predictions and the 
Chapman et al. sources is about 0.25 in alpha, and the discrepancy is greater 
for the OS predictions. In order for even the smaller of the discrepancies to be 
explained by the correlation in Figure~\ref{alpha-s60-both}, 
$\alpha^{850}_{1.4}$ would 
have to decrease by approximately 0.25 over a luminosity range of a factor of $\simeq$4.
This seems unlikely looking at the slow change in $\alpha$, but we cannot exclude the 
possibility.}

\item{One possibility could be that the Chapman et al.  sources have significant AGN 
activity contributing to their 1.4 GHz emission (an extreme example in the local 
Universe would be NGC\,6166 and the other six or so outliers in the OS sample 
FIR--radio correlation shown in 
Figure~\ref{lum-radio-lor}) --- if dust is heated to higher temperatures in 
stronger radiation fields, e.g. near to an AGN, we could expect to see both higher 
1.4 GHz emission and higher dust temperature. However, it seems unlikely that 
the radio and FIR emission from these galaxies is dominated 
by radiation from an active nucleus because a) the galaxies are not generally strong 
X-ray sources (Waskett et al. 2003), b) the radio morphologies are not those 
typical of radio-loud AGN (Chapman et al. 2004), and c) the optical spectra are 
often those of starbursts rather than AGN (Chapman et al. 2005). In addition, we find
that excluding the 18 submm galaxies which exhibit AGN characteristics does not affect the 
result we see in Figure~\ref{high-red}.}

\item{Another possibility, which would contribute to the scatter of the Chapman et al. 
sources in Figure~\ref{high-red}, is that some of the redshifts are unreliable. In order 
to test this we selected a subsample of the Chapman et al. sources with the most robustly 
determined redshifts --- following Aretxaga 
et al. (2007), those whose 
spectroscopic redshifts are derived from the identification of at least two spectral 
features and which have unambiguous optical/IR/radio counterparts. We find that using 
this reduced sample does not affect the result we see in Figure~\ref{high-red}, and as 
such it seems unlikely that unreliable redshifts are the explanation.}

\item{The fourth possibility, which we feel is most likely, is that the relationship between 
FIR and radio emission at high redshift is different from that at low redshift. 
Given the very different conditions in the Universe ten billion years ago compared 
to today, it would actually be quite surprising if the relationship were the same. 
Moreover, there is already evidence that the radio 
sources associated with high-redshift dust sources are quite different from those today. 
Chapman et al. (2004) find that the 
majority of the high-redshift sources have larger radio sizes ($\sim10$\,kpc) 
than are found for local Ultra-Luminous Infrared Galaxies (ULIRGs, $\lesssim1$\,kpc). 
This would suggest that the physical conditions may be different for 
low and high redshift sources, in which case one might expect to see some evolution 
of the FIR--radio relation.}

\end{enumerate}

\section{Discussion}

Although the explanation of the relation between radio and FIR emission at low redshift 
is still uncertain (see Section~\ref{intro}), there are still some general 
statements one can make about the possibility of 
this relationship evolving with redshift. 

We will assume that the standard model is correct: the 
FIR emission is from dust heated by young OB stars and the radio emission is 
from relativistic electrons which were produced by supernovae. Within the 
standard model (which invokes an explicit dependence on star formation), there are 
two explanations 
of the slope (of roughly unity on a log-log plot) of the low-redshift relationship. 
The first is the calorimeter model (V\"olk 1989). In this model, the electrons 
radiate all their energy before they 
escape from the galaxy; variations in the magnetic field change the rate at which an 
individual electron loses its energy but do not change the FIR--radio ratio. The 
second type of explanation is that the slope, which is not precisely unity, 
is a cosmic conspiracy. In models of 
this type, electrons do escape from galaxies before they radiate all their energy, which 
would lead to a lower FIR--radio ratio in galaxies with stronger magnetic fields. 
However, this is avoided in the models by fine-tuning them so that, for example, 
electrons diffuse more rapidly out of galaxies 
in which the magnetic fields are stronger (e.g. Chi \& Wolfendale 1990). 
Other fine-tuning models include, for example, those of Helou \& Bicay (1993) 
and Niklas \& Beck (1997). 
The calorimeter model seems 
the more natural explanation, but its disadvantage has been that the 
radio spectra of galaxies do not 
show the clear down-turn at high frequencies expected for an aging population, 
although Thompson et al. (2006) have argued that this is the result of another 
cosmic conspiracy -- of ionization and 
bremsstrahlung losses flattening the spectra. 

If the calorimeter model is correct, we would probably 
expect a change in the FIR--radio ratio with redshift in the opposite direction 
to the one we actually observe. In this model, increases in the magnetic field in 
high-redshift galaxies, which one might expect as the result of increased 
turbulence in the interstellar medium, will have no affect on 
the FIR--radio ratio. The one effect, however, that one might expect is that 
the electrons will lose a greater fraction of their energy to inverse 
Compton losses, because of the increase in the energy 
density of the microwave background and because of the strong interstellar 
radiation field in these galaxies. In this case, however, we would expect 
the FIR--radio ratio to increase with redshift because an electron is losing 
more of its energy by inverse Compton losses than by synchrotron losses. 

The cosmic conspiracy models contain more free parameters and therefore lead 
to less definite 
predictions. However, since the high-redshift galaxies 
detected in the deep submm surveys do appear to have larger radio sizes than 
the ULIRGs found in the local Universe, one might expect the relativistic 
electrons to escape more slowly from the 
galaxies. In this case, one would expect the FIR--radio ratio to decrease 
with redshift. This is in qualitative agreement with what we observe. Therefore, 
our results provide some evidence against the calorimeter model.

One important question is whether the SLUGS SEDs are suitable templates for the 
high-redshift objects in the Chapman et al. (2005) sample. Unfortunately, for these distant 
submm sources there are very limited observations at wavelengths which probe the 
peak in the dust SED and the spectral slope in the radio regime. The current best 
estimates of the FIR SEDs come from SHARC-2 observations at 350\micron\/ (probing rest 
frame 120\micron\/ at $z=2$) presented by Kov\'acs et al. (2006) and Coppin et al. (in 
preparation). This allows a basic estimate of the dust temperature independent 
of assumptions about the FIR-radio correlation, and these authors find an average 
dust temperature of $\sim 35$ K for submm selected galaxies --- very similar to the 
average temperature for the SLUGS IRS sample (Dunne et al. 2000b). In this respect, the 
FIR/submm part of the IRS SLUGS SEDs are probably a sensible choice of template. In 
the radio, 
however, it is less obvious that things could be the same at both low and high redshift. 
For example, there are very 
few measurements of the radio spectral index for high-redshift submm sources. A 
flattening of the average spectral index compared to that for the low-redshift sources 
($\alpha=0.7$) would tend to move the high-redshift sources in the direction observed in 
Figure~\ref{high-red}. Hunt, Bianchi \& 
Maiolino (2005) notice that SEDs appropriate for compact dwarf galaxies (which have a higher 
component of free-free emission and so flatter spectral indices) produce better 
photometric redshift estimators than SEDs based on more typical IR luminous galaxies. 
The reason why the high-redshift sources are generally brighter at radio wavelengths 
compared to our SLUGS $\alpha^{850}_{1.4}$--z predictions 
may be uncovered 
by deep multi-wavelength radio observations of submm sources which will investigate 
their radio properties, as well as better coverage of the IR SED (e.g. from Herschel).

\section{Conclusions}

Using the results of the OS SLUGS sample we have investigated the relationships between 
FIR--submm and radio properties for `normal' local galaxies, compared to bright 
\IRAS\/ galaxies (using the IRS SLUGS sample from Dunne et al. 2000b). The results for 
the OS galaxies have then been used to assess the reliability of the Carilli \& Yun 
(1999, 2000) radio--submm redshift estimator technique, for the first time using a 
sample containing significant fractions of cold dust. 

We find the following results.
\begin{itemize}
\item{There is a very tight correlation between 1.4\,GHz luminosity and 60-\micron\/ 
luminosity for the OS sample galaxies. There is much more scatter in the correlation 
between radio and 850\mic luminosity, thus providing evidence that the standard explanation 
for the FIR--radio luminosity correlation is correct and that it is the population of 
young (OB) stars which both heats the dust and provides the source of radio emission. }

\item{Using the SEDs of the 17 OS galaxies detected at both 850- and 450-\mic we have 
shown that for a sample of `normal' galaxies there is much more scatter in the 
$\alpha^{850}_{1.4}$-redshift relation than seen for the 25 most luminous \IRAS\/ 
galaxies of the IRS SLUGS sample, most likely since the OS objects contain large 
fractions of cold dust. From this we conclude that in order for the Carilli \& Yun 
radio--submm redshift indicator method to be reliable as a redshift estimator for deep 
submm sources one would need measurements of the temperature of the dust in those objects.} 

\item{We have compared our OS $\alpha^{850}_{1.4}$-redshift relations to the deep submm 
observations of Chapman et al. (2005) and demonstrate that the large majority of these deep 
SCUBA sources must have very different properties to our sample of low-redshift SLUGS 
galaxies. We find that the majority of the deep SCUBA galaxies are brighter at radio 
wavelengths than any of our predictions. We cannot exclude the possibility that this
simply reflects that our low-redshift galaxies are not suitable templates for the high-redshift
SCUBA galaxies, but we think it more likely that this shows the FIR--radio relation is 
evolving with redshift.}
\end{itemize}

\section*{Acknowledgments}
The authors would like to thank Rob Ivison and Itziar Aretxaga for useful discussions and 
the anonymous referee for helpful comments.


\begin{thebibliography}{}
\bibitem{} Appleton P.N. et al., 2004, ApJS, 154, 147
\bibitem{} Aretxaga I., et al., 2007, MNRAS, preprint (astro-ph/0702503)
\bibitem{} Becker R.H., White R.L., Helfand D.J., 1995, ApJ, 450, 559
\bibitem{} Bettens R.P.A., Brown R.D., Cragg D.M., Dickinson C.J., Godfrey P.D., 1993, 
MNRAS, 263, 93
\bibitem{} Bicay M.D., Helou G., 1990, ApJ, 362, 59
\bibitem{} Blain A.W., 1999, MNRAS, 309, 955
\bibitem{} Boulanger F., P\'erault M, 1988, ApJ, 330, 964
\bibitem{} Carilli C.L., Yun M.S., 1999, MNRAS, 513, L13
\bibitem{} Carilli C.L., Yun M.S., 2000, AJ, 530, 618
\bibitem{} Chapman S.C., Smail I., Windhorst R., Muxlow T., Ivison R.J., 2004, ApJ, 611, 732
\bibitem{} Chapman S.C., Blain A.W., Smail I., Ivison R.J., 2005, AJ, 622, 772
\bibitem{} Chi X., Wolfendale A.W., 1990, MNRAS, 245, 101
\bibitem{} Condon J.J., 1992, ARA\&A, 30, 575
\bibitem{} Condon J.J., Helou G., Sanders D.B., Soifer B.T., 1990, ApJS, 73, 359
\bibitem{} Condon J.J., Cotton W.D., Greisen E.W., Yin Q.F., Perley R.A., Taylor G.B., 
Broderick J.J., 1998, AJ, 115, 1693
\bibitem{} Cox M.J., Eales S.A.E., Alexander P., Fitt A.J., 1988, MNRAS, 235, 1227
\bibitem{} Devereux N.A., Eales S.A., 1989, AJ, 340, 708
\bibitem{} Dunne L., Eales S.A., 2001, MNRAS, 327, 697
\bibitem{} Dunne L., Clements D.L., Eales S.A., 2000a, MNRAS, 319, 813
\bibitem{} Dunne L., Eales S., Edmunds M., Ivison R., Alexander P., Clements D.L., 2000b, 
MNRAS, 315, 115
\bibitem{} Eales S.A., Lilly S.J., Gear W.K., Dunne L., Bond R.J., Hammer F., 
Le F\`evre O., Crampton D., 1999, ApJ, 515, 518
\bibitem{} Fitt A.J., Howarth N.A., Alexander P., Lasenby A.N., 1992, MNRAS, 255, 146
\bibitem{} Garrett M.A., 2002, A\&A, 384, 19
\bibitem{} Groves B.A., Cho J., Dopita M., Lazarian A., 2003, PASA, 20, 252
\bibitem{} Helou G., Bicay M.D., 1993, ApJ, 415, 93
\bibitem{} Helou G., Soifer B.T., Rowan-Robinson M., 1985, ApJ, 298, L7
\bibitem{} Hoernes P., Berkhuijsen E.M., Xu C., 1998, A\&A, 334, 57
\bibitem{} Hughes D.H. et al., 1998, Nat, 394, 241
\bibitem{} Hunt L., Bianchi S., Maiolino R., 2005, A\&A, 434, 849
\bibitem{} Kov\'acs A., Chapman S.C., Dowell C.D., Blain A.W., Ivison R.J., Smail I., 
Phillips T.G., 2006, ApJ, 650, 592
\bibitem{} Lewis G.F., Chapman S.C., Helou G., 2005, AJ, 621, 32
\bibitem{} Lisenfeld U., V\"olk H.J., Xu C., 1996, A\&A, 306, 677
\bibitem{} Longair M.S., 1983, C\&T, 99, 32
\bibitem{} Murgia M., Helfer T.T., Ekers R., Blitz L., Moscadelli L., Wong T., 
Paladino R., 2005, A\&A, 437, 389
\bibitem{} Murphy E.J. et al., 2006, ApJ, 638, 157
\bibitem{} Niklas S., Beck R., 1997, A\&A, 320, 54
\bibitem{} Pierini D., Popescu C.C., Tuffs R.J., V\"olk H.J., 2003, A\&A, 409, 907
\bibitem{} Pohl M., 1994, A\&A, 287, 453
\bibitem{} Price R., Duric N., 1992, ApJ, 401, 81
\bibitem{} Roussel H., Helou G., Beck R., Condon J.J., Bosma A., Matthews K., 
Jarrett T.H., 2003, ApJ, 593, 733
\bibitem{} Smail I., Ivison R.J., Blain A.W., 1997, ApJ, 490, L5
\bibitem{} Thompson T.A., Quataert E., Waxman E., Murray N., Martin C.L., 2006, ApJ, 
645, 186
\bibitem{} Vlahakis C., Dunne L., Eales S., 2005, MNRAS, 364, 1253
\bibitem{} V\"olk H.J., 1989, A\&A, 218, 67
\bibitem{} Waskett T.J. et al., 2003, MNRAS, 341, 1217
\bibitem{} Yun M.S., Reddy N.A., Condon J.J., 2001, ApJ, 554, 803
\end{thebibliography}
\end{document}